# Probing the presence and absence of metal-fullerene electron transfer reactions in helium nanodroplets by deflection measurements

John W. Niman,[a]  Benjamin S. Kamerin,[a]  Thomas H. Villers,[a]

Thomas M. Linker,[b]  Aiichiro Nakano,[b]  Vitaly V. Kresin*[a]

[a] *Department of Physics and Astronomy, University of Southern California,
Los Angeles, CA 90089-0484, USA*

[b] *Collaboratory for Advanced Computing and Simulations,
University of Southern California, Los Angeles, CA 90089-0242, USA*

## ABSTRACT

Metal-fullerene compounds are characterized by significant electron transfer to the fullerene cage, giving rise to an electric dipole moment. We use the method of electrostatic beam deflection to verify whether such reactions take place within superfluid helium nanodroplets between an embedded $C_{60}$ molecule and either alkali (heliophobic) or rare-earth (heliophilic) atoms. The two cases lead to distinctly different outcomes: $C_{60}Na_n$ ($n$=1-4) display no discernable dipole moment, while $C_{60}Yb$ is strongly polar. This suggests that the fullerene and small alkali clusters fail to form a charge-transfer bond in the helium matrix despite their strong van der Waals attraction. The $C_{60}Yb$ dipole moment, on the other hand, is in agreement with the value expected for an ionic complex.

# 1 INTRODUCTION

Since the discovery of alkali-doped fullerides and their superconductivity,[1,2] it has been known that metal atom-fullerene structures with significant charge transfer are readily formed in the bulk, surface,[3] and gas phases.[4–6] Gas-phase studies are informative because they permit a molecular-level analysis of the charge transfer process.

The formation of an ion pair implies the appearance of a large dipole moment, and indeed studies of fullerene-alkali systems by molecular beam electric deflection[7] have demonstrated high electric susceptibilities related to extremely large (~10–20 D) dipole moments in these systems. Interestingly, these experiments showed that at higher temperatures the alkali metal atoms and clusters appear to skate about the surface of the fullerene. Analogously, complexes of $C_{60}$ with transition metals were found to have dipole moments of 6-10 D.[8]

Helium nanodroplet embedding offers a method to assemble and study such systems at very low temperatures and within a superfluid environment.[9] Rotational and vibrational motion are greatly suppressed at the nanodroplet temperature of 0.37 K, and polar complexes can be strongly oriented by an external electric field and interrogated by pendular spectroscopy[10,11] and electrostatic deflection.[12]

However, there is a hurdle: alkali atoms are strongly heliophobic and are not wetted by the helium medium.[13] As a result, alkali atoms and small clusters do not submerge into helium nanodroplets and instead reside in surface dimples.[14–17] On the other hand, recent work by Renzler *et al.*[18] provided evidence, based on electron ionization mass spectrometry, that co-doping helium nanodroplets with the highly polarizable $C_{60}$ molecule induces Na atoms, as well as small clusters of Na and Cs, to submerge into the nanodroplet. It was suggested that this could derive from the strong long-range van der Waals attraction between the partners,[19] or from electron transfer via a harpoon reaction. A follow-up computational study[20] suggested that harpoon-type charge transfer does take place between $Cs_2$ and fullerene dopants.

The presence or absence of an ion pair within a nanodroplet can be directly established by the electrostatic deflection method. In this work we apply it to the sodium-$C_{60}$ system and demonstrate that for $C_{60}Na_{n=1-4}$ there is in fact no evidence of a strong electric dipole appearing in the system, and hence of the partners approaching closely enough to form an ionic bond.

For a comparative measurement, we performed a deflection measurement for nanodroplets doped with $C_{60}$ and ytterbium atoms. Yb was chosen due to the combination of a favorable ionization energy and vapor pressure. It is also pertinent that ytterbium-intercalated fulleride conductors have been synthesized[21] and were found to exhibit superconductivity. While to the best of our knowledge Yb previously has not been used as a nanodroplet dopant, it is unambiguously expected to solvate, analogously to Eu, another rare-earth atom.[22] Here we observe a sizable deflection indicative of the formation of a large electric dipole. The magnitude of the dipole is in very good agreement with the computed dipole moment for a Yb atom positioned on a pentagonal face of the fullerene, implying that an electron is readily transferred and a bound $C_{60}Yb$ system is formed.

## 2 EXPERIMENT

As described in refs. 12,23 a supersonic beam of helium nanodroplets is generated by expansion of ultrahigh purity grade helium gas at 80 bar stagnation pressure through a 5 micron nozzle held at a temperature of 15 K. The beam is skimmed, and chopped by a rotating wheel; the beam velocity is measured to be $\approx$375 m/s. It then passes through two heated stainless steel pick-up cells, with the first containing $C_{60}$ powder (99.9%) and the second a small lump of Na (99.99%, loaded under hexane to combat oxidation) or of Yb chips (99.95%). The dopants are picked up by sequential collisions with the droplet beam and their thermal energy is promptly dissipated by partial evaporation of the droplet.[9]

The pick-up process follows Poisson statistics. The cell temperatures were stabilized to optimize the signal to a desired average number of dopants. The temperature of $C_{60}$ was fixed to a value between 370˚ C and 380˚ C. This yielded sufficient intensity in the mass spectrum at the single $C_{60}^+$ mass while keeping the intensity of the $C_{60}$ dimer to a negligible level. (At this temperature many of the fullerene's vibrational modes[24] are activated, hence pick-up of one molecule results in the nanodroplet shrinking by ~9000 He atoms.) Deflection measurements of $C_{60}Na_n$ were taken with the metal-containing cell temperature ranging from 190˚ C to 210˚ C, and of $C_{60}Yb$ with this cell at 360˚ C.

The beam then travels to the deflection chamber where it is collimated by a 0.25 mm × 1.25 mm slit, and passes between two 15 cm-long high voltage electrodes which create an electric field and a collinear field gradient directed perpendicular to the beam axis. With an applied voltage of 20 kV, a field strength of 82 kV/cm with a gradient of 338 kV/cm$^2$ is achieved.[25] The nanodroplets then traverse a 1.25 m free-flight region and enter the aperture of a quadrupole mass spectrometer (Ardara Technologies), where they are ionized by electron impact with an electron energy of 70 eV and detected using a pulse-counting channeltron multiplier with a digital counter system synchronized to the chopper.[26] With this arrangement, detection profiles can be acquired even with counting rates as low as a few per second.

Among the peaks in the mass spectrum there are those which correspond to bare $C_{60}M^+$ ions, where M is a metal atom or cluster. These ions are ejected from the droplet following charge exchange with a $He^+$ hole generated by electron-impact ionization.[15] Note that they may derive either from post-ionization encounters between $C_{60}^+$ and M (or between $C_{60}$ and $M^+$), or from the ionization of bound $C_{60}M$ complexes preformed in the nanodroplets. The aim of the measurement is to ascertain whether nanodroplets contain any polar complexes of the latter type, formed by a charge transfer reaction. It is worth reiterating that the deflection step occurs before ionization of the beam, and therefore the reactants and the complexes are electrically neutral.

Deflections profiles were collected by setting the mass spectrometer to these peaks and translating the detection chamber on a precision linear slide, controlled by a stepper motor, under two conditions: "field-off" and "field-on." The former establishes the original cross section of the beam, and the latter determines the magnitude of beam deviation under the influence of the electric field. As mentioned above, the very low temperature of the helium nanodroplets allows for nearly complete orientation of the permanent dipole of the embedded dopant along the electric field. This permits the force from the electric field gradient on the dipole to become so strong as to deflect the entire heavy nanodroplet on the scale of millimeters.

The initial nanodroplet size distribution in the beam is calibrated by deflections using a dopant with a known dipole moment, in this case CsI. For the conditions employed in the present measurements, the mean size was found to be $\approx 2 \times 10^4$ helium atoms per droplet. Using this information and the "field-off" profile as input, we perform a Monte Carlo simulation of the deflection, as detailed in our previous publications,[12,23,27,28] accounting for droplet shrinkage by evaporation and for the pick-up and ionization probabilities, and solve for the dipole moment of the embedded complex by fitting the simulated "field-on" profile to the data.

## 3 RESULTS

### 3.1 Sodium

Fig. 1 displays the $C_{60}Na_n$ mass spectra over the temperature range used for deflection measurements. The peak widths are partially due to the presence of $^{13}C$ isotopes in the fullerene. Note that the ion signals are quite weak compared to the intensity of the bare fullerene peak. The $C_{60}Na^+$ signal is the weakest of all, including at cell temperatures outside of the range shown. Thus, sodium has a low propensity for forming bound complexes with $C_{60}$. To determine whether they ever establish direct contact as neutral dopants, we look for evidence of the formation of a dipole moment.

Electrostatic beam deflections were performed at the temperatures shown in Fig. 1. At these settings a given mass peak acquires sufficient intensity for a profile measurement, while the larger ones remain weak, minimizing the likelihood of their fragmentation contaminating the peak of interest. This strategy could not be adopted for $C_{60}Na$ due to its low peak intensity but was followed for $C_{60}Na_{2-4}$.

Fig. 2 shows the undeflected and deflected beam profiles acquired with the mass spectrometer set to the $C_{60}Na_n^+$ peaks. For all measured $n$ the deflection is essentially negligible within the accuracy of the measurement. This is in striking contrast with the 14–16 D dipole moment of the ionic $C_{60}Na$ molecule,[8,29] which would have resulted in deflections on the order of several millimeters,[30] and confirms that no charge-transfer bound complexes between the fullerene cluster and the sodium atom form within the nanodroplet.

### 3.2 Ytterbium

The case for fullerenes and Yb atoms is qualitatively different. This was already hinted at by the mass spectra, where the ratio of the $C_{60}Yb^+$:$C_{60}^+$ peak intensities was an order of magnitude higher than for $C_{60}Na_n^+$/$C_{60}^+$. Deflection measurements made the distinction apparent. Fig. 3(a) shows the "field-off" and "field-on" profiles of the $C_{60}Yb$ mass peak. A very sizable deviation is immediately evident, in contrast with the results for sodium, and establishes that the embedded complex has formed a strong permanent electric dipole.

The fitting procedure described in Sec. 2 allows us to quantify the magnitude of this dipole moment.[32] Fig. 3(b) displays the result of the optimized fit and demonstrates good agreement with the experimental data. Based on this procedure we assign a dipole moment of 8 ± 1.5 D to the $C_{60}Yb$ complex. The estimated error margin derives partially from the data fit and partially from the uncertainty in the branching ratio involved in the charge transfer from $He^+$ to the dopant.[12] The result unambiguously demonstrates that the rare earth metal atom contacts and donates charge to the fullerene.

As shown in the next section, the experimentally deduced value of the dipole moment is in good agreement with a calculation of the $C_{60}Yb$ complex, in particular for Yb atoms positioned at the pentagonal face of the fullerene. According to the computation, this situation has a marginally advantageous binding energy (although it is known that systems embedded in helium nanodroplets are not always able to reach their lowest energy geometries). Measurements of crystalline fullerene-ytterbium compounds suggest that the pentagonal face indeed is preferred for charge transfer,[33,34] supporting our results.

## 4 CALCULATIONS

To further validate the result for $C_{60}Yb$, the dipole moments and binding energies of free-space molecules were computed by plane-wave based Density Functional Theory (DFT). Calculations were performed in the Vienna Ab-Initio Software Package (VASP).[35,36] Electronic states were computed within the frozen core approximation using the projected augmented wave-vector (PAW) method[37,38] with projectors generated for the C $2s$ and $2p$ states and the Yb $6s$ and $5p$ states. The strongly correlated Yb $f$ electrons were assumed not to be involved in bonding and were kept within the core of the pseudopotential. A plane wave cut-off energy of 500 eV and the Perdew–Burke–Ernzerhof (PBE)[39] styled generalized gradient approximation (GGA) for the exchange-correlation functional were used. The $C_{60}$ and $C_{60}Yb$ molecules were placed in the center of a non-cubic 15 Å × 18 Å × 20 Å box to remove spurious image interactions. Optimization was first performed on the $C_{60}$ molecule to obtain its ground state structure. For computation of the binding energies and dipole moment the ground state $C_{60}$ structure was kept frozen during the optimization of $C_{60}Yb$. Linear dipole corrections were used to correct the errors introduced by the periodic boundary conditions.[40]

The results of these calculations are presented in Table 1 alongside similar ones for the dipole moment of isolated $C_{60}Na$. The latter are in reasonable agreement with previous computations[29] and partially validate the present results for $C_{60}Yb$. As mentioned in the preceding section, it is found that settling the Yb atom on the pentagonal face yields a marginally higher binding energy and a noticeably lower dipole moment.

**Table 1**  DFT calculations for the dipole moment and binding energy of a Yb or Na atom optimized on either the hexagonal or pentagonal face of the $C_{60}$ fullerene.

|  | Dipole moment (D) | | Binding energy (eV) | |
| --- | --- | --- | --- | --- |
|  | Pentagon | Hexagon | Pentagon | Hexagon |
| $C_{60}Yb$ | 8.5 | 13.0 | 0.58 | 0.54 |
| $C_{60}Na$ | 13.4 | 12.8 | 1.16 | 1.21 |

## 5 SUMMARY

The experiments detect no sizable electric dipole moment appearing when a fullerene molecule, followed by between one and several sodium atoms, are embedded in helium nanodroplets. Therefore even if a heliophobic alkali atom is pulled inside the nanodroplet by the presence of $C_{60}$,[18] it appears that they continue to be separated by a layer of helium and neither short-range electron transfer nor a longer-range harpoon reaction take place.

As for the absence of an electric dipole moment for $C_{60}Na_{2-4}$, one can envision two scenarios. One is that the sequentially picked up atoms assemble into a small sodium cluster on or near the droplet surface but still fail to approach the fullerene within the droplet sufficiently to transfer charge and form a bond. This contrasts with $C_{60}Na_n$ agglomerates forming in neat molecular beams.[41] An alternative possibility is that multiple sodium atoms do attach to the fullerene but arrange themselves in symmetric configurations, as calculated for lowest-energy structures due to Coulomb repulsion between positive sodium ions.[29] This cannot be excluded, but would require either a sequence of individual atom-$C_{60}$ agglomeration events (which appear unlikely within the nanodroplets in view of the data), or attachment of a Na cluster followed by its separation into individual atoms and their subsequent rearrangement around the cage (which, however, would need to proceed in the very low-temperature nanodroplet environment). Consequently, the data do not support the theoretical picture[20] of an alkali dimer undergoing a harpoon reaction with $C_{60}$ and settling into an ionic arrangement with the latter.

In contrast to sodium, we observe that a very strong permanent dipole moment is formed between ytterbium atoms (which are wetted by helium) and $C_{60}$, revealing successful electron transfer and bond formation in this system. Comparison with modeling of the $C_{60}Yb$ molecule suggests that the ytterbium atom prefers to locate above the pentagonal face of the fullerene. Interestingly, while the harpoon reaction is suppressed in binary collisions in the gas phase[19] due to unfavorable Franck-Condon factors,[42] in the present case the strong reactive channel is kept open thanks to removal of the accompanying vibrational excitation by the helium matrix.

It would be interesting to extend such measurements to larger alkali clusters, because above a certain critical size they begin to submerge into the nanodroplet by themselves.[43,44] This should promote charge transfer and dipole formation, analogous to observations on $C_{60}$-alkali cluster complexes in free space.[7] Interesting complementary information also could be derived from spectroscopic experiments, since near-IR absorption peaks of the fullerenes have been shown[45] to be sensitive to the oxidation state of $C_{60}{}^{n-}$.

### ACKNOWLEDGEMENTS

We would like to express our appreciation to Prof. G. Benedek for his eminent contributions to research on helium nanodroplets, fullerenes, and nanoclusters. We would like to thank L. Kranabetter and P. Scheier for productive discussions and R. E. Pedder for advice on sensitivity calibration of the quadrupole mass spectrometer. The research of J.W.N, B. S. K., T. H. V. and V. V. K. was supported by the U. S. National Science Foundation, Division of Chemistry. T. M. L. and A. N. were supported by the National Science Foundation, Award OAC 2118061. Simulations were performed at the Center for Advanced Research Computing of the University of Southern California. The figures for this article were generated using the SciDraw scientific figure preparation system.[46]

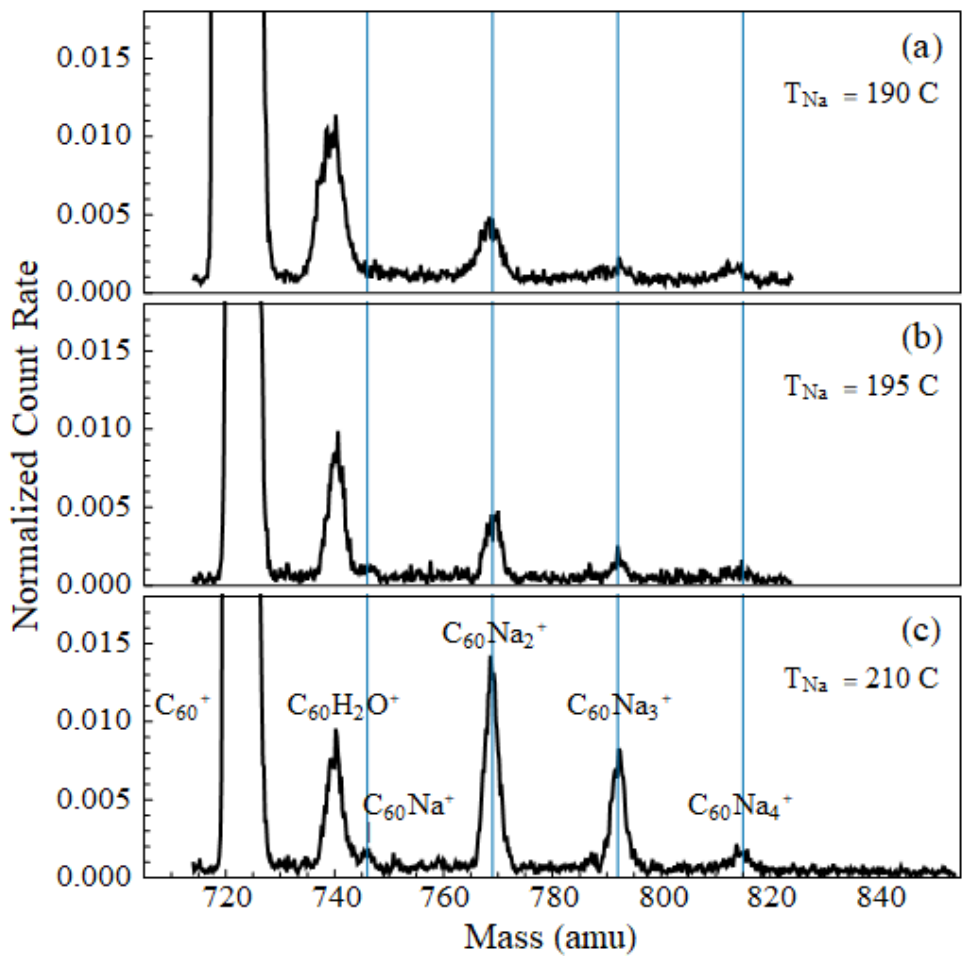

**Fig. 1** Mass spectra of $C_{60}Na_n^+$ complexes with the corresponding Na pick-up cell temperatures indicated. All spectra are normalized to the intensity of the $C_{60}^+$ peak. The $C_{60}Na^+$ signal is weak for all temperatures. Panels (a) and (b) show the conditions used for deflection measurements of the $C_{60}Na_2^+$ and $C_{60}Na_3^+$ peaks, respectively, while (c) gives the conditions for $C_{60}Na^+$ and $C_{60}Na_4^+$ deflections.

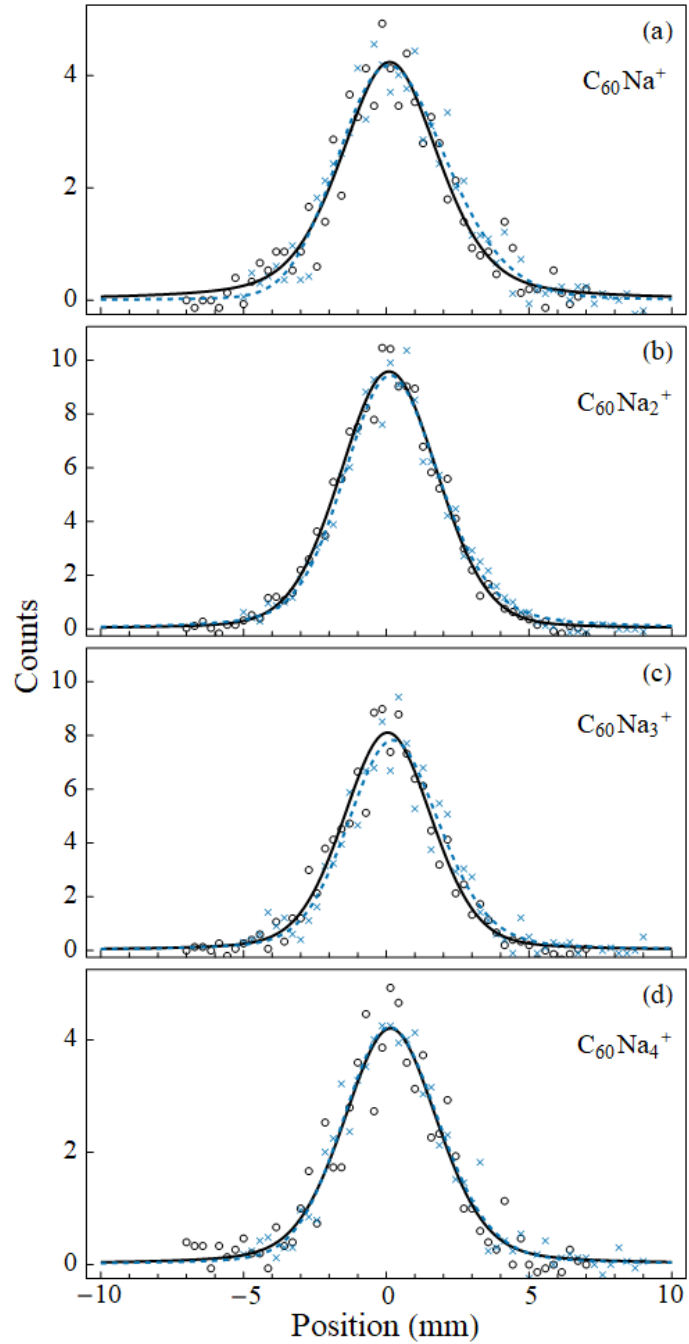

**Fig. 2** Beam deflection measurements for $C_{60}Na_{n=1-4}$. Circles (crosses) denote data points with the electric field turned off (on). The counting rates of selected ions in the strongly collimated beam were on the order of a few per second. The solid lines are symmetric pseudo-Voigt[31] fits to the "field-off" profile , while the dashed lines are asymmetric pseudo-Voigt fits to the "field-on" profile. In all cases the beam deflection is negligible.

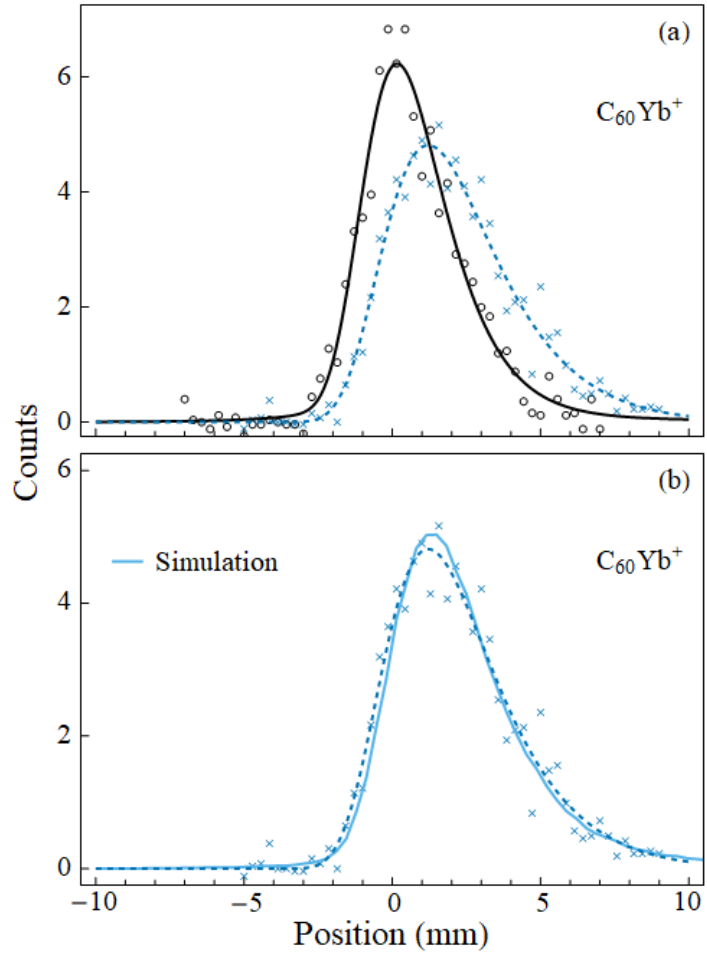

**Fig. 3** Panel (a) shows the beam deflection measurements of $C_{60}Yb$ with the same notation as Fig. 2. A slight offset of the second pick-up cell added minor skewness to the profiles, hence in this figure both the "field-off" and the "field-on" ones were fitted to an asymmetric pseudo-Voigt function. The shift of the profile centroid is $\approx 1.4$ mm. Panel (b) shows the same "field-on" profile (dashed line) and data points, together with the Monte Carlo simulation (solid line) for the optimized dipole moment value of 8.5 D.